\documentclass[aps,prl,preprint,showpacs,superscriptaddress,notitlepage]{revtex4-2}

\usepackage{hyperref}

\usepackage{graphicx}
\usepackage{natbib}
\usepackage{wrapfig}
\newcommand{\be}{\begin{equation}}
\newcommand{\ee}{\end{equation}}
\newcommand{\nn}{\mbox{} \nonumber \\ \mbox{}}
\newcommand{\ba}{\begin{eqnarray}}
\newcommand{\ea}{\end{eqnarray}}
\newcommand{\om}{\omega}
\newcommand{\Alfven}{Alfv\'{e}n\,}

\newcommand{\Bf}{{magnetic field}}
\newcommand{\Bfs}{{magnetic fields}}

\newcommand{\Lf}{Lorentz factor}

\newcommand\eg{{\it{e.g.,}}}

\newcommand\lo{\mathrel{\raise.3ex\hbox{$<$}\mkern-14mu\lower0.6ex\hbox{$\sim$}}}
\newcommand\go{\mathrel{\raise.3ex\hbox{$>$}\mkern-14mu\lower0.6ex\hbox{$\sim$}}}

\usepackage{amsmath}
\usepackage[normalem]{ulem}
\usepackage{color}

\begin{document}


\title{Chaotic cosmic ray acceleration  in nonlinear  Alfven waves}
\author{Maxim Lyutikov}
\affiliation{
Department of Physics and Astronomy, Purdue University, \\
 525 Northwestern Avenue,
West Lafayette, IN
47907-2036 }

\begin{abstract}
We demonstrate that  motion of  a   particle in a single coherent  linearly polarized (LP) \Alfven  wave with relative  amplitude $\delta = B_w /B_0  \gtrsim 0.25 $ becomes chaotic.
  As a result,  a  sufficiently strong, near resonant  \Alfven waves ($r_L k_\parallel \sim 1$, $r_L $ is relativistic cyclotron radius, $k_\parallel$ is wavevector)   leads to transient trapping and  phase mixing, imitating  classical scattering. 
Evolution of pitch angle  via Alfven chaos is non-diffusive - a particle on trapped trajectory can reverse its velocity in one scattering event. This both    avoids the problem of the  90$^\circ$ barrier, and   increases  the  acceleration rate  for those special particle.
 \end{abstract}

\maketitle 

 \newpage 
 

\section{Introduction}

Acceleration of  cosmic rays (CRs) is an important astrophysical problem \citep{1949PhRv...75.1169F,1987PhR...154....1B,1978MNRAS.182..147B,1991SSRv...58..259J,1998ApJ...493..694M,2009ApJ...694..626R,2013A&ARv..21...70B}.  At play is the  non-linear balance between self-generated plasma turbulence and  accelerated CRs \citep{1969ApJ...156..445K,1975MNRAS.172..557S,1983RPPh...46..973D,1994A&A...281..220A,2004MNRAS.353..550B,2006MNRAS.371.1251A,2011MNRAS.410...39B,2013SSRv..178..201B}. 

The dominant model is the Diffusive Shock Acceleration (DSA) whereby small pitch angle scattering on resonant \Alfven waves leads to slow, diffusive  evolution of the pitch angle, and eventual back-scattering
\citep{1966ApJ...146..480J,1999ApJ...520..204G}.  

Several problems of DSA include, first, slow acceleration rate \cite{1983A&A...125..249L,2001JPhG...27.1589K,2003A&A...403....1P}.   Second is the   "$90^\circ$ barrier" problem   \citep{1971ApJ...169...41K}:  in standard quasi-linear theory, the pitch-angle diffusion coefficient $D_{\alpha\alpha}$ vanishes at $\alpha = 90^\circ$. As we demonstrate, the present model does not suffer from the barrier problem.

Nonlinear effects in wave-particle interaction were always a suspect in resolving difficulties encountered by the quasilinear theories 
\citep{1972SvPhU..14..549Z,2003ApJ...590L..53M,2006ApJ...642..244M,2009ASSL..362.....S,2019ApJ...873...13R,2017PhRvE..95c3207B}.

It is generally understood that particle motion in nonlinear \Alfven waves can become chaotic 
\citep{1992JGR....9713853K,2001PhPl....8.4713C,2002PhRvL..89B1102Y,2005JASTP..67.1852M,2010ApJ...720..503C,2010ApJ...720..503C,2023ApJ...951...88W,1990PhFlB...2..606K}.   Below, we discuss a surprisingly  simple set-up for chaos onset. 
We demonstrate that trajectories of   particles, relativistic or not,  interacting with  single, monochromatic, linearly polarized Alfv\'en wave,  propagating along a guide magnetic field, become  chaotic for wave amplitudes $\delta \equiv B_w/B_0 \gtrsim 0.25$.  No turbulent cascade with many modes, no wave-wave interactions, no parametric instabilities are required---just  a high energy   particle and one harmonic nonlinear   \Alfven wave. 

Mathematically, our approach is closest to  Ref.  \citep{1990PhFlB...2..606K}. Conceptually, in Ref. \cite{2006ApJ...642..244M}  trapping of CRs in nonlinear waves were discussed ({\eg} their Fig. 4).

\section{CR motion in static Alfven  wave} 

\subsection{Particle  motion via Hamilton-Jacobi approach} 

Alfven  waves in the Solar wind and the   interstellar medium have  subluminal velocities,  much smaller than the speed of light.
\be
\beta _A \equiv \frac{v_A}{c} =  \frac{ \om_{B,i} }{\om_{p,i}} \ll 1 
\ee
where $ \om_{B,i}  $ and  $\om_{p,i}$ correspondingly cyclotron, and plasma frequencies associated with  plasma protons. 
We can then transform to the Alfven frame, where electromagnetic fields are static, and the electric field is zero.  In that frame motion of a particle is in a purely magnetic configuration, consisting of a guide field $B_0$ and fluctuating field $B_w$. In the Alfven frame  the energy of a particle is conserved. 

It is natural then to consider particle motion using Hamilton-Jacobi (HJ) approach. 
For a relativistic particle with rest mass $m$ and charge $q$, the  Hamiltonian ($\Phi = 0$) is:
\be
H = c\sqrt{m^2c^2 + (\mathbf{P} - q\mathbf{A})^2} 
\ee
where  $\mathbf{P} $ is generalized momentum and $\mathbf{A}$ is vector potential. 

There is a choice  freedom of vector potential $\mathbf{A}$. Depending on the polarization of the wave, circular or linear (CP-LP), it is more convenient to choose the vector potential in the symmetric-cylindrical gauge for CP, or Landau gauge for LP. Hamiltonian particle dynamics in linear and circularly polarized waves is very different. In CP there is an extra conserved quantity,  due to helical symmetry,  that makes dynamics integrable. 

Here we're concerned with the non-integrable motion of a particle in a strong nonlinear  {\it linearly polarized}  \Alfven  wave.  It is then more convenient to choose the vector potential in the Landau gauge.  For magnetic field along the $x$ direction, the wave propagating along $x$  as well,  with polarization along $z$ direction we can choose  (in $x-y-z$ coordinates) 
 \be
 \mathbf{A} = \left(0, - z + \frac{\delta }{k}\sin(k x), 0\right) B_0 
 \ee
The  relativistic  Hamiltonian is 
\be
H = c\sqrt{m^2c^2 + P_x^2 + P_z^2 + \left( P_y + q B_0 z - \frac{q \delta B_0}{k} \sin(k x) \right)^2} 
\ee
where $k$ is the wave vector of the  wave vector in the \Alfven  frame; it is slightly different by  Lorentz transformation with $\beta_ A \ll 1$ from the wave vector in the lab frame.

Since the Hamiltonian has no explicit time and $y$  dependence, for the action S  we use the ansatz 
\be
S(\mathbf{r}, t) = W(\mathbf{r}) - Et + P_y y
\ee
 where $E$ is the total relativistic energy. The resulting  relativistic Hamilton-Jacobi equation:
\be
 \left(\frac{\partial W}{\partial x}\right)^2 + \left(\frac{\partial W}{\partial z}\right)^2 + \left( P_y + q B_0 z - \frac{q \delta B_0}{k} \sin(k x) \right)^2 = \frac{E^2}{c^2} - m^2c^2 
 \ee

Applying Hamilton's canonical equations  we find (energy is conserved $\gamma = {H}/{mc^2}=$ constant)
\ba && 
\dot{x} = \frac{c^2 P_x}{H} = \frac{P_x}{\gamma m} 
\nn  && 
 \dot{y} = \frac{c^2}{H}\left( P_y + q B_0 z - \frac{q \delta B_0}{k} \sin(k x) \right) 
 \nn && 
\dot{z} = \frac{c^2 P_z}{H} = \frac{P_z}{\gamma m} 
\nn && 
\dot{P}_x =
 q \dot{y} \delta B_0 \cos(k x) 
\nn && 
\dot{P}_z =
-q \dot{y} B_0 
\ea

The corresponding  accelerations are 
\ba && 
\ddot{x} =  \dot{y} \delta \Omega_c  \cos(k x) 
 \nn &&
 \ddot{y} = (\dot{z} - \dot{x} \delta \cos(k x))  \Omega_c
 \nn  &&
\ddot{z} = - \dot{y} \Omega_c
\nn &&
\Omega_c = \frac{q B_0}{\gamma m c}
\label{Omegac} 
\ea

We can isolate $y$ dynamics
\be
 \dot{y} = \Omega_c  z -  \delta   \frac{ \Omega_c } {k} \sin(k x) + C
 \ee
 where $C$ is constant integration. It can be set to zero  by a proper boost along $y$. 
 
 The dynamic equations become
\ba &&
\ddot{x} =  \left(  \delta  z \cos(k x) - \frac{\delta ^2}{2k} \sin(2k x) \right) \Omega_c^2 
\nn &&
\ddot{z} + \Omega_c^2 z = \frac{\delta }{k} \sin(k x)   \Omega_c^2 
  \label{main0} 
\ea

Renormalizing time by  $\Omega_c$ and coordinates  by $k$, the dimensionless set of equations governing motion of a particle in guide field and linearly polarized \Alfven wave, in the \Alfven frame, becomes
\ba &&
\ddot{x}  = \delta  \cos (x) (z- \delta  \sin (x))
\nn &&
\ddot{z}+z  = 2
   \delta  \sin (x) \cos (x)
   \label{main} 
   \ea
 Since  constant relativistic factors are absorbed into definition of  $\Omega_c$, this equation is valid for non-relativistic  particles as well. 
 In (\ref{main} linear terms on the lhs correspond to motion with constant  velocity along $x$ and cyclotron oscillations in $z$.


\subsection{Chaos onset }

 This system  (\ref{main})  exhibits deterministic chaos depending on the strength of the relative  wave amplitude $\delta$.  It  has  a 4-dimensional phase space, but only one constant of motion (energy). Hence it does not satisfy the  Liouville-Arnol'd condition  for integrability \citep{arnold1989mathematical}.  The trigonometric  functions  introduce strong, non-separable nonlinearities that  couple the parallel and perpendicular dynamics.

The transition from regular, orderly motion to chaos in this type of wave-particle interaction follows the {Chirikov resonance overlap mechanism}  \citep{Chirikov1979}.  In the absence of the  wave, $\delta  =0$,  $\ddot{x}=0$. The particle moves at a  constant velocity along the $x$-axis while executing  harmonic Larmor gyration in the perpendicular plane. The system is fully integrable.
For a weak wave perturbation $\delta \ll 1$, Kolmogorov-Arnold-Moser (KAM) invariant tori structurally restrict the phase space; the particle's trajectory is perturbed but remains  bounded, and quasi-periodic.
 In the {chaotic regime ($\delta \gtrsim  \delta _{crit} \sim  1$),   the primary and secondary nonlinear resonances between the particle's spatial bounce motion in the wave potential and its fundamental gyro-motion begin to overlap. This resonance overlap destroys the protective KAM surfaces, leading to  stochasticity.
  
  This behavior is illustrated in  Fig. \ref{poincare_maps} where we show Poincare maps for particle motion in static guide \Bf\ plus linearly polarized static wave.  Figure \ref{poincare_maps}  also  illustrate  a transition from the regular KAM  trajectories   at $\delta \ll 1$ to chaotic at   $\delta \to 1$.  The transition to chaos occurs near  $\delta _{crit}  \sim 0.2$.

 \begin{figure}
 \includegraphics[width=0.95\textwidth]{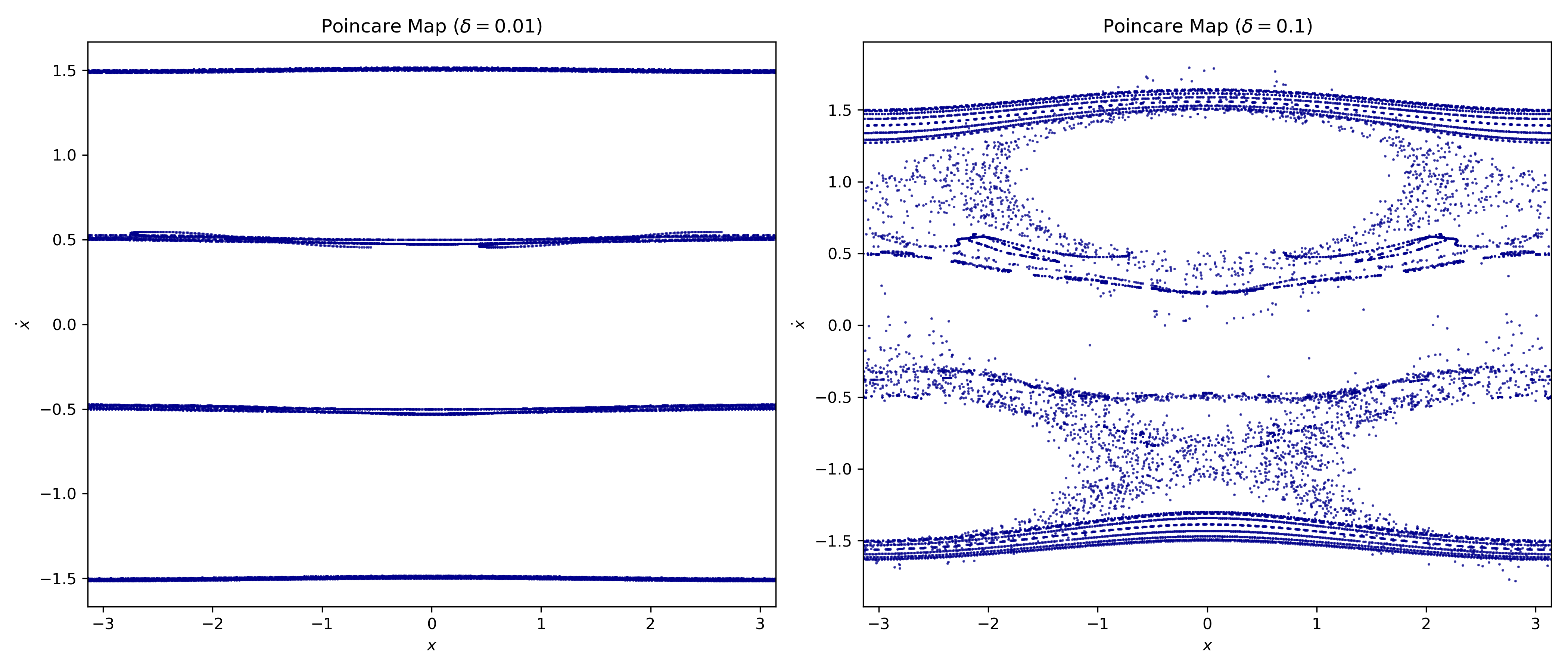}
  \includegraphics[width=0.95\textwidth]{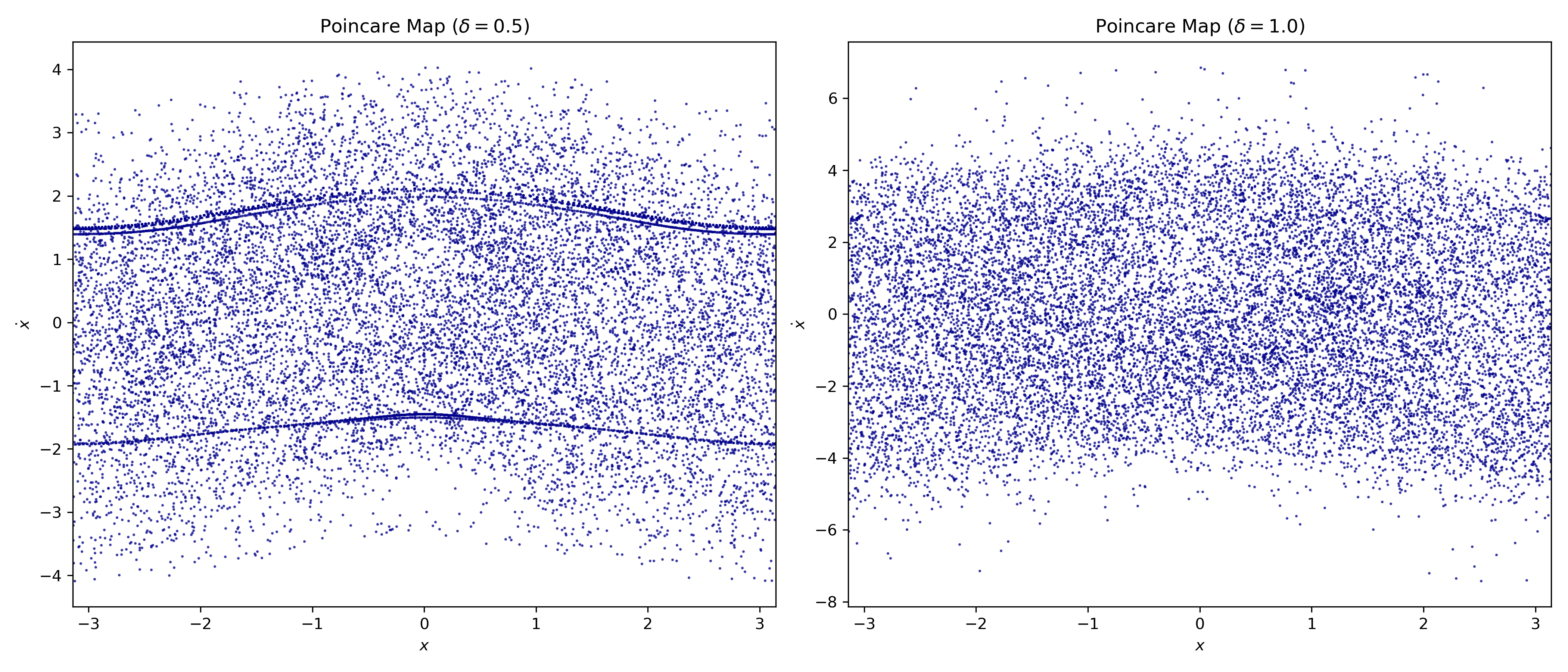}
  \caption{  Poincare maps (phase space in static  magnetic field. As the amplitude of the fluctuating field increases, particle motion becomes chaotic for  $\delta \gtrsim  \delta _{crit}  \approx 0.2$.}
\label{poincare_maps}
\end{figure}



To characterize  timescale of the instability, 
we computed the Lyapunov exponent $\lambda(\delta)$, Fig. \ref{Lyapunov}. At small $\delta \leq \delta _{crit}  $ the Lyapunov exponent  is zero, jumping to finite values above $\delta _{crit}  \approx 0.25$.

 \begin{figure}
 \includegraphics[width=0.95\textwidth]{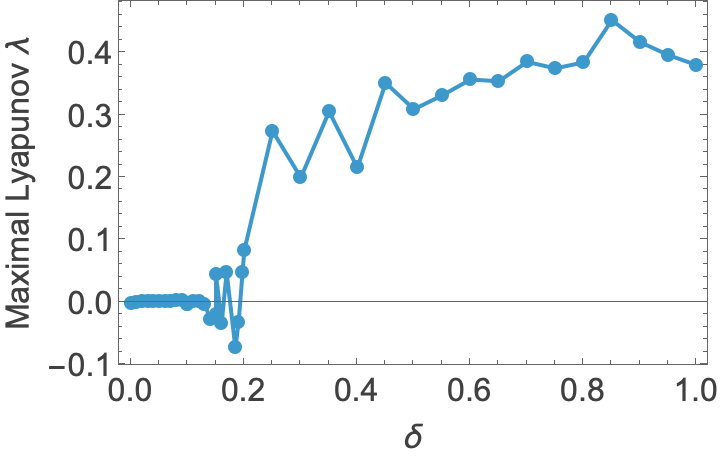}
  \caption{Lyapunov exponent as function of nonlinearity  $\delta$. At $\delta \to 0 $, the Lyapunov exponent  is effectively zero -  there is no chaos. Starting some critical $\delta_{crit} \sim 0.25$, the  Lyapunov exponent  jumps to non-zero values, indicating  the onset of a chaos. (Relatively large fluctuations in $\lambda$ are due to sampling of the perturbed trajectories used to calculate $\lambda$.)}
\label{Lyapunov}
\end{figure}

The  more extended computations to large $\delta$  shows that the maximal Lyapunov exponent does not return to zero at large $\delta$,  but saturates at $\lambda \sim 0.4$. 
For the static-field system, chaos persists at all  $\delta$; 
 there is no return to regularity at large $\delta$, even though in the limit $\delta \to \infty$, a particle in a LP wave, the motion is integrable. \citep[this resembles particle dynamics in wave and transverse \Bf][]{2026arXiv260704359L}.
 
 Since time in (\ref{main}) is normalized to relativistic cyclotron frequency, 
in dimensional units the constant dimensionless Lyapunov exponent $\lambda$  translates to  
\be
\lambda_{\rm phys} =  \lambda  \Omega_c \propto 1/\gamma
\label{lambda_phys}
\ee
Sufficintly high above the resonance, Lyapunov exponent  decreases and the motion returns to deterministic.

Since $\lambda$ saturates at $\approx 0.4$,  the  physical Lyapunov exponent at resonance ($\gamma = \gamma_{\rm res}$, $kr_L = 1$)  is 
\be
\lambda_{\rm phys} \approx 0.4\, \Omega_c  \approx 0.4\, k_\parallel c
\label{lambda_res}
\ee


\subsection{ Small  $\delta$ limit}

To understand the onset of chaos, we expand system (\ref{main})    in the perturbation parameter $\delta$:
\ba && 
\ddot{x} \approx \delta z \cos(x) \label{eq:approx_x} 
\nn && 
\ddot{z} + z \approx 0 \label{eq:approx_z}
\label{smalldelta}
\ea

For  the unperturbed perpendicular solution $z(t) = a \sin(t ) $,  the parallel perturbation equation \eqref{eq:approx_x} give 
\begin{equation}
\ddot{x} = \delta a \sin(t ) \cos(x) =\frac{\delta a}{2} \left[ \sin(x + t ) - \sin(x - t ) \right]
\end{equation}

Resonances occur when the phase of a driving term matches the particle's parallel motion.  This isolates two primary resonance velocities in the phase space:
\begin{equation}
v_x =  \pm 1
\end{equation}
The absolute distance separating the centers of these two adjacent resonances is:
$
\delta v =  2
$
(recall that we normalized space by $k$ and time by $\Omega_c$, hence velocity by $\Omega_c/k$).

Near the $v_x = 1$ resonance, the phase dynamics are dominated by the second wave component. We introduce a slow resonant phase variable defined by $\theta = x - t$. Differentiating $\theta$ twice with respect to time yields\ a nonlinear pendulum equation:
\begin{equation}
\ddot{\theta} + \frac{\delta a}{2} \sin(\theta ) = 0
\end{equation}

The total energy of this effective island pendulum system is:
\begin{equation}
E_{\text{island}} = \frac{1}{2}\dot{\theta}^2 - \frac{\delta a}{2} \cos(\theta )
\end{equation}
The boundary separating trapped bounded motion (libration) from untrapped running motion (rotation) is the separatrix, defined by the energy value $E_{\text{island}} = \frac{\delta a}{2}$. Substituting this value yields the max velocity boundary:
\begin{equation}
\frac{1}{2}\dot{\theta}_{\max}^2 - \frac{\delta a}{2} \cos(\theta ) = \frac{\delta a}{2} \implies \dot{\theta}_{\max} = \pm 2\sqrt{\frac{\delta a}{2}}\sin\left(\frac{\theta }{2}\right)
\end{equation}

The maximum half-width $\Delta v$ of the resonance island occurs at the peak velocity separation on the separatrix, where the sine term equals unity:
\begin{equation}
\Delta v = 2 \sqrt{\frac{\delta a}{2}} = \sqrt{2 \delta a}
\end{equation}
Due to symmetry, both primary resonances ($v_x = \pm 1$) share an identical half-width: $\Delta v_1 = \Delta v_2 = \sqrt{2\delta a}$.

According to the Chirikov criterion, the threshold for the transition to global stochasticity is reached when the sum of the adjacent resonance half-widths equals or exceeds the absolute distance separating their center points:
\begin{equation}
\Delta v_1 + \Delta v_2 \ge \delta v
\end{equation}
The final criterion for the onset of global chaos:
\begin{equation}
\delta \ge \frac{1}{2a}
\label{deltacrit}
\end{equation}

To map the resonance zones over time, we  integrate (\ref{smalldelta}) and construct a Poincare surface of section by plotting the parallel phase space coordinates \((x, \dot{x})\) every time the perpendicular coordinate \(z\) crosses a specific plane (e.g., \(z = 0\) with \(\dot{z} > 0\)), Fig. \ref{Poincare1}

\begin{figure}
 \includegraphics[width=0.95\textwidth]{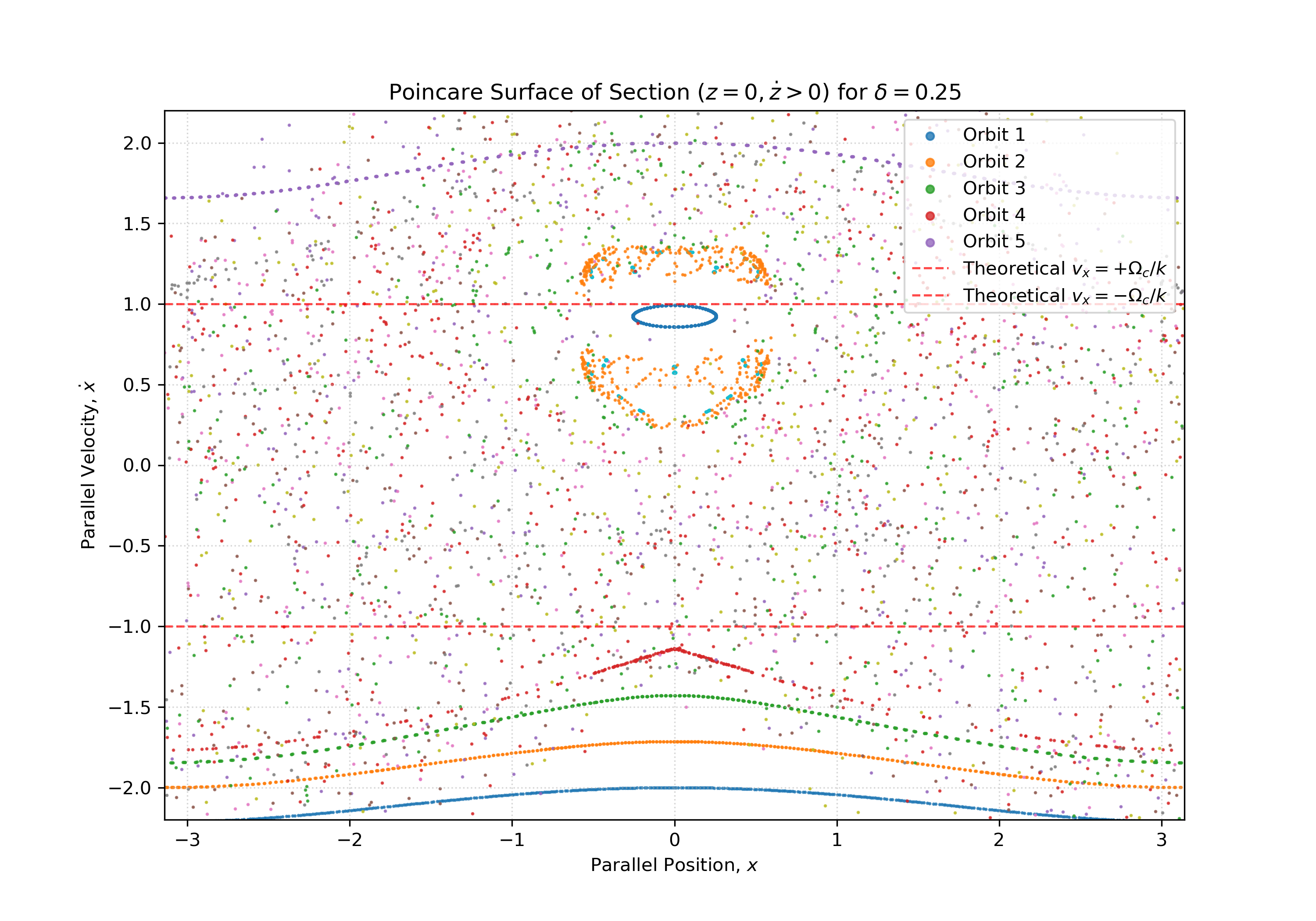}
  \caption{Poincare map for system  (\ref{main}), $\delta =0.5$, $a=1.5$, Chirikov threshold  (\ref{deltacrit})  $\delta_{crit}  = 0.333$,  periodic conditions in $x$, integration over $3000/(2\pi)$ periods. This map has higher value  of energy $a$ than  in Fig. \ref{poincare_maps}  and scans over  initial coordinate $-\pi < x_0 < \pi$; this  populates different invariant tori  than  in Fig. \ref{poincare_maps}. Upper-lower asymmetry   is due to $\delta^2$ term in  (\ref{main}).}
\label{Poincare1}
\end{figure}

The smooth lines trace out stable island chains, while scattered points reveal the chaotic sea.   The orbits passing near the boundary lines of these large islands  appear as a loose scatter of randomized points rather than smooth continuous loops. This marks the onset of the local stochastic layer.  At this intermediate perturbation strength (\(\delta = 0.25\)), continuous invariant horizontal lines  still span across the domain near \(\dot{x} \approx 0\), demonstrating that global chaos is not yet active.

To illustrate particle trapping, in Fig. \ref{trapping} we plot trapping time for an \Alfven  Gaussian pulse of dimensionless length $6$ depending on the initial velocity. The plot  reveals alternating parameter windows of smooth, instantaneous transmission interspersed with intervals of long-term trapping. 
\begin{figure}
 \includegraphics[width=0.95\textwidth]{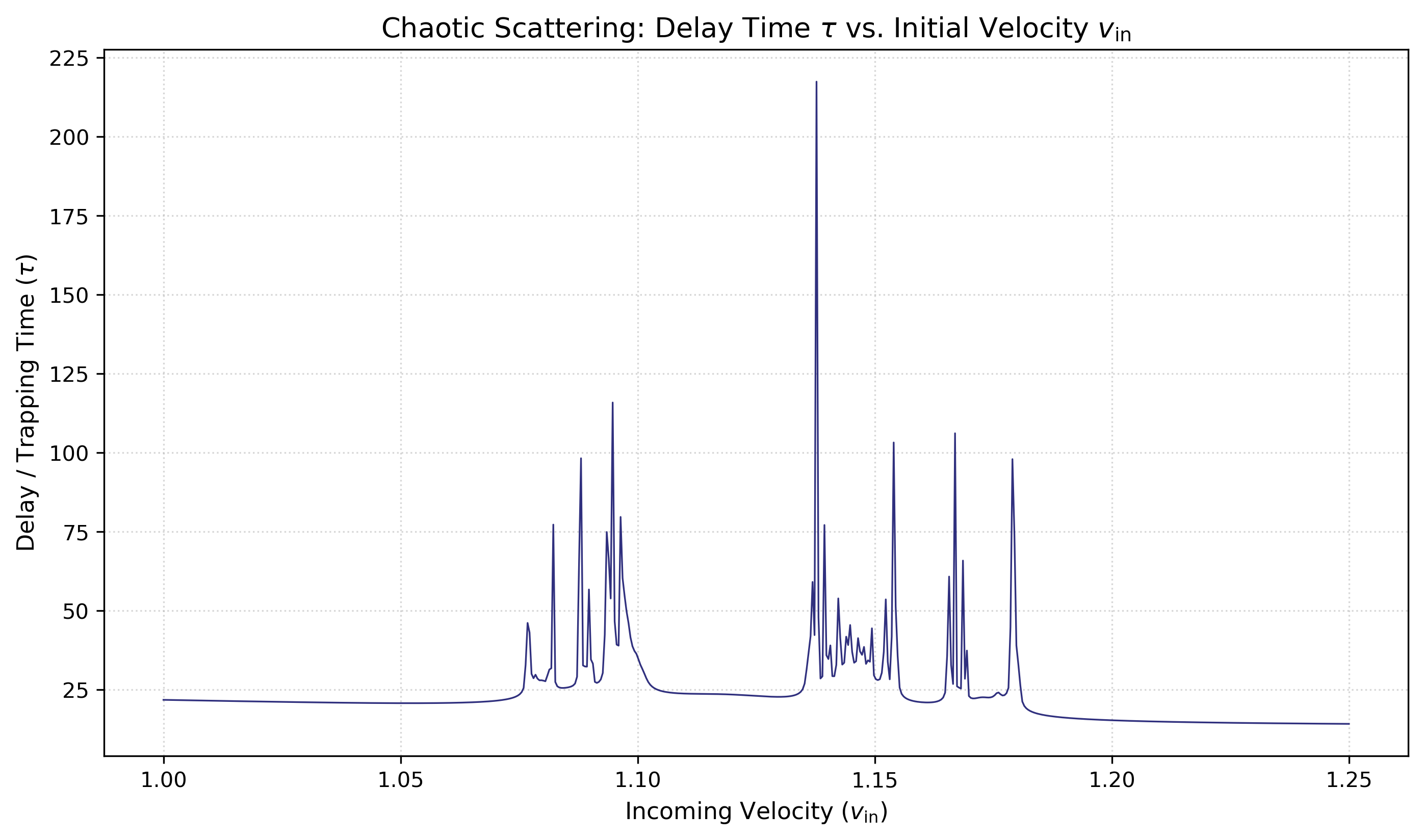}
  \caption{Trapping time for an \Alfven  Gaussian pulse of dimensionless length $6$ depending on the initial velocity. Parameters are $\delta =0.4$, $a = 1.5 $. Velocity is normalized to resonant $\Omega_c/k$. Overall shift   of $\sim 20$ is just the straight fly-by regime. Trapped trajectories are spikes of long residence time of a particle within the wave packet.}
\label{trapping}
\end{figure}

Fig. \ref{trapping} illustrates that trapping (or  reversal) time is a highly non-smooth function of the normalized velocity. In the vicinity of the resonant regime  $v\sim 1$ there are regions of long trapping. If a particle hits one of those trajectories, it becomes trapped and escapes nearly in random direction. The evolution of pitch angle in that case is non-diffusive.


\section{Discussion} 

 The onset of global Hamiltonian chaos for large-amplitude ($\delta \gtrsim 1$) Alfv\'en waves has  consequences for the transport and acceleration of cosmic rays in astrophysical plasmas.  
The central result of this work is that a  {single monochromatic, linearly polarized Alfv\'en wave} propagating along a guide magnetic field produces Hamiltonian chaos for wave amplitudes $\delta \equiv B_w/B_0 \gtrsim 0.25$.  

This chaos onset is different from stochastic instability in turbulent \Bfs,  random walk of magnetic field line, \citep{1972SvPhU..14..549Z,1972SvPhU..14..549Z,2009AdSpR..43.1429S}. In our case chaos onset occurs in a single, harmonic, mildly nonlinear wave.

 The standard theory of CR pitch-angle scattering (diffusive shock acceleration, DCA) \citep{1969ApJ...156..445K,1975MNRAS.172..557S,1983RPPh...46..973D} operates in the weak-wave limit $\delta \ll 1$, quasi-linear cosmic ray transport.  Resonances are isolated, perturbatively narrow, and never overlap.  The theory predicts regular diffusion, not Hamiltonian chaos, because it cannot access the nonlinear resonance overlap regime. Earlier studies of chaos in Alfv\'en waves \citep{1990PhFlB...2.2581H,2007NPGeo..14...17C} investigated chaotic behavior in the  {wave evolution equations} (driven derivative nonlinear Schr\"odinger equation), i.e., wave--wave chaos.  These are fundamentally different from the single-particle Hamiltonian chaos in a fixed wave field described here.

 The onset of chaos  can be equivalently interpreted as a {chaotic scattering process}.  
 An   incoming particle is mapped  to a highly randomized, uncorrelated outgoing state after passing through a localized interaction zone. When an incoming, untrapped particle enters this stochastic layer, it  is no longer merely transmitted or reflected directly. Instead, it  undergoes a long sequence of irregular, transient loops inside the localized wave zone. Two incoming trajectories with  close initial velocities will separate exponentially fast inside the chaotic  interaction zone. They will execute  different numbers of loops before being ejected, mapping to vastly different exit parameters. During the period of transient trapping, the particle's memory of its initial entry state is  erased via  phase mixing. When the particle finally escapes the interaction zone, its exiting phase is statistically randomized. Macroscopically, this collisionless interaction mimics the appearance of a classical scattering.

The model also offers resolution of the "$90^\circ$ barrier".  In standard quasi-linear theory of cosmic ray transport, the pitch-angle diffusion coefficient $D_{\alpha\alpha}$ vanishes at $\alpha = 90^\circ$ ($v_\parallel = 0$) because the resonant wavenumber $k_{\rm res} = \omega_c/(\gamma v_\parallel)$ diverges---there is no wave to resonate with at zero parallel velocity.  The non-linear dynamical resonance discussed here does not suffer from this limitation.

The present model seems to be in agreement with modern PIC simulations:  in Ref. \citep{2025ApJ...988..144H},  upstream of the shock,  no  broad spectrum of  waves was observed. Instead,   localized, large-amplitude waves (SLAMS)  were numerically observed.  Test-particle simulations using such fields seem to have mimicked  scattering and DSA results.

The interaction of intense electromagnetic waves with charged particles is well studied in the context of laser-driven acceleration, where wave amplitudes routinely reach $a_0 \gg 1$.  
However,  laboratory \Bfs\ are not high enough to achieve cyclotron resonance with the laser pulse. 

Large-amplitude Alfv\'en waves ($\delta \gtrsim 1$) are indeed observed in astrophysical environments: in the solar wind \citep{1985ApJ...299..122S}, near collisionless shocks, and in the self-generated turbulence upstream of supernova remnant blast waves.  The chaos threshold $\delta \gtrsim 1$ is therefore astrophysically  relevant;  the resulting pitch-angle stochasticity may  also provide the mechanism for crossing the $90^\circ$ barrier that has long challenged quasi-linear transport theory.

On the other hand, stability of nonlinear LP \Alfven waves has been questioned \citep{1991PhFlB...3.1407M}. Resulting nonlinear steepening may instead lead to  a periodic train of Alfven shocks. 

Importantly, in treating  the particle dynamics we neglected ponderomotive effects which would make the problem time-dependent. It is expected then variable \Lf\ may induce additional resonance crossing. In addition, effects of anomalous cyclotron resonance   \citep[e.g.,][]{1960SPhUs...2..874G, 1996PhyU...39..973G} may become important. \citep[We also mention here  the importance of the  anomalous cyclotron resonance for pulsar coherent emission][]{1979SvAL....5..238M,1991MNRAS.253..377K,1999MNRAS.305..338L,1999ApJ...512..804L}.

Finally, we comment that the chaos onset may be important for ion heating in the Solar wind \citep{2010ApJ...720..503C}, and 
a somewhat related result - chaos onset in non-linear X-mode propagating perpendicular to the \Bf, \citep{2026arXiv260704359L}.

We would like to thank  Elena Amato, Mikhial Malkov,  and Anatoly Spitkovsky for comments on the manuscript.

\bibliographystyle{apsrev}
\bibliography{Alfvenchaos.bib}

\end{document}